\documentclass[useAMS,usenatbib]{mn2e}

\usepackage{amsmath}
\usepackage{epsfig}

\title[Narrow-gap etalons]{General Formulation for the Calibration and Characterization of Narrow-gap Etalons: the OSIRIS/GTC Tunable Filters Case}
\author[J.J. Gonz\'alez et al.]{J.J. Gonz\'alez$^{1}$, J. Cepa$^{2,3}$
J.I. Gonz\'alez-Serrano$^{4}$\thanks{E-mail:
gserrano@ifca.unican.es}, and M. S\'anchez-Portal$^{5}$
\\
$^{1}$Instituto de Astronom\'\i a, Universidad Aut\'onoma de M\'exico, Apartado Postal 70-264, 04510, M\'exico D.F, M\'exico\\
$^{2}$Instituto de Astrof\'\i sica de Canarias, E-38205, La Laguna, Tenerife, Spain\\
$^{3}$Departamento de Astrof\'\i sica, Universidad de La Laguna, E-38206, La Laguna, Tenerife, Spain\\
$^{4}$Instituto de F\'\i sica de Cantabria (CSIC - Universidad de Cantabria), E-39005, Santander, Spain\\
$^{5}$Herschel Science Centre, ISDEFE/ESAC, E-28691, Villanueva de la Ca\~nada, Madrid, Spain
}
\begin{document}

\date{Accepted xxxx. Received xxxx ; in original form xxxx}

\pagerange{\pageref{firstpage}--\pageref{lastpage}} \pubyear{2014}

\maketitle

\label{firstpage}

\begin{abstract}
Tunable filters are a powerful way of implementing narrow-band imaging mode over wide wavelength ranges, 
without the need of purchasing a large number of narrow-band filters covering all strong emission or absorption 
lines at any redshift. However, one of its main features is a wavelength variation across the field of view, 
sometimes termed the Òphase effectÓ. In this work, an anomalous phase effect is reported and characterized for the OSIRIS instrument at the 10.4m Gran Telescopio Canarias. The transmitted wavelength across the field of view of the instrument depends, not only on the distance to the optical centre, but on wavelength. This effect is calibrated for the red tunable filter of OSIRIS by measuring both normal-incidence light at laboratory and spectral lamps at the telescope at non-normal incidence. 

This effect can be explained by taking into account the inner coatings of the etalon. In a high spectral resolution etalon, the gap between plates is much larger than the thickness of the inner reflective coatings. In the case of a tunable filter, like that in OSIRIS, the coatings thickness could be of the order of the cavity, which changes drastically the effective gap of the etalon. We show that by including   thick and dispersive coatings into the interference equations, the observed anomalous phase effect can be perfectly reproduced. In fact, we find that, for the OSIRIS red TF, a two-coatings model fits the data with a rms of 0.5\AA\ at all wavelengths and incidence angles. This is a general physical model that can be applied to other  tunable-filter instruments.

\end{abstract}


\section{Introduction}

OSIRIS (Optical System for Imaging and low Resolution Integrated Spectroscopy) is an imager and low-resolution optical spectrograph for the segmented 10.4m GTC telescope. The main characteristic that differentiates it from similar instruments for 8-10m telescopes, is the narrow band imaging mode implemented using two tunable filters (TF), over a circular field of view (FOV) of 8 arcmin in diameter. The blue TF covers from 3650 up to 6700\AA\ while the red one covers from 6500 up to 9350\AA. The full widths at half maximum (FWHM) that are available at each wavelength range from 8 to 20\AA, depending on wavelength. Both, wavelength and FWHM can be tuned with accuracy of $\sim 1$\AA\ in few tens of milliseconds \citep{b9}.

TFs are a flexible approach for narrow band imaging in large telescopes, without the need of purchasing a large set of filters for the study of different emission or absorption lines at any redshift (see e.g. \citealt{b2}). 
They have demonstrated advantages with respect to standard narrow-band filters \citep{b20} and also over low-resolution spectroscopy. TFs allow for relatively high-resolution measurements ($R\sim 500$) compared to low-resolution grisms, and can be tuned to avoid telluric lines. TFs have been proved to be much more efficient for observing absorption features in, for instance, atmospheres of extrasolar planets (\citealt{b28}, \citealt{b27}).

The first mention to tunable filters applied to astronomy in the literature, seems to be that of \citet{b21}. While initially applied to solar studies \citep{b22} due to stability reasons, \citet{b1} proposed a piezo-electric tuning with servo-control driven tunable filter device, with parallelism stability enough to be used in space payload or Cassegrain foci of telescopes. Its routine application to galactic and extragalactic astronomy had to wait till the advent of the Taurus Tunable Filter for the Anglo-Australian Telescope (AAT) and for the William Herschel Telescope (WHT) \citep{b2}. For a review of scientific results with these telescopes see \citet{b23}. Since then, tunable filters have been in use in several telescopes, including the Maryland-Magellan \citep{b15}, the SOAR Telescope \citep{b24}, and is planned for the JWST \citep{b25}, among others. Since its first scientific operation in March 2009, more than 100 refereed papers using  OSIRIS at the GTC, have been published or are accepted for publication. Also, OTELO \citep{b26}, the deepest emission line survey to date, has just finished its first pointing, demonstrating the possibility to routinely achieving 5$\sigma$ depths better than 10$^{-18}$ erg cm$^{-2}$ s$^{-1}$ at a FWHM of 1.2nm using the TF of OSIRIS at GTC.

However, TFs have several limitations, among which the most conspicuous one is the variation of the wavelength across the FOV, symmetrically from the optical centre of the instrument outwards, due to its insertion into a collimated beam, that has been sometimes called a Òphase effectÓ (see \citealt{b3} for a review).

Several other instruments have used or are currently using TFs. For example, the Taurus Tunable Filter \citep{b2}, currently decommissioned, was at operation at the 3.9m AAT. The wavelength variation across its 10 arcmin field of view was of 15\AA\ at the rest frame H$\alpha$ line. The Robert Stobie Spectrograph \citep{b12} for the 11m Southern African Large Telescope, has a circular FOV of 8 arcmin, with a wavelength variation across the field of 21\AA, also at the rest frame H$\alpha$ line. Finally, the Maryland-Magellan Tunable Filter (MMTF, \citealt{b15}) for the 6.5m Magellan-Baade telescope, has a larger field of view (27 arcmin in diameter) with a larger phase effect. OSIRIS at the GTC, although with a smaller field of view, has a large phase effect as well, of $\sim 80$\AA. It is clear that obtaining the maximum scientific outcome from OSIRIS TFs requires a proper characterization of the wavelength variation across the FOV.

The scientific operations of GTC, with OSIRIS as the only instrument then available, started in March 2009, before the end of the commissioning of the red TF. This decision was taken with the aim of speeding up the scientific exploitation of the telescope. However, several months later, it was reported that the phase effect was significantly different from that theoretically expected, and stated in the operation manual of the OSIRIS TFs. 
As a consequence, a systematic calibration data gathering campaign started, with the collaboration of GRANTECAN, the company in charge of operating GTC, with the aim of calibrating the real wavelength dependence over the OSIRIS red TF FOV. 

The structure of the paper is as follows. First, the basic theoretical approximate equations (as usually found in many text books) are presented, and the theoretical wavelength dependence derived. In section 3, previous reports on the anomalous phase effect, together with its impact on astronomical observations using TFs are discussed. In Section 4 the general equations are derived. Section 5 is devoted to the real characterization of TFs, Section 6 to the calibration techniques, Section 7 to the results obtained for OSIRIS TF and, finally, the general conclusions are presented.

\section[]{Basic zero order equations for a Fabry-Perot}
\subsection[]{Filter transmission and orders of interference}

The equations of this section are easy to derive, and can be found on many books about optics (as for example, \citealt{b8}), 
and are given as an introduction so that the reader can be familiar with the terms and nomenclature used along this work. 

Typical etalons on astronomical telescopes are designed with the same dielectric multilayer coating on both reflective surfaces of the cavity, and are usually operated under collimated (or very slow) beams. The general equation for the intensity transmission coefficient of an ideal (perfectly flat plates used in a parallel beam) Fabry-Perot (FP) as a function of wavelength is

\begin{equation}
\tau_r=\left(\frac{T}{1-R}\right)^2\left[1+\frac{4R}{(1-R)^2}\sin^2\left(\frac{2\pi\mu d\cos\theta}{\lambda}\right)\right]^{-1}
\end{equation}
 where $T$ is the transmission coefficient of each coating (plate-cavity boundary), $R$ is the mean reflection coefficient, $d$ is the plate separation (plate spacing or ÒgapÓ), $\mu$ is the refractive index of the medium in the cavity (usually air, $\mu=1$) and $\theta$ is the angle of incident light. Constructive interference occurs when the optical path difference (OPD) in the gap medium and the OPD due to the reflective coatings add up to be a multiple of a given wavelength $\lambda$. Considering the incidence angle $\theta$, relative to the normal of the parallel plates, a zero-order approximation for this interference condition in such a cavity is,

\begin{equation}
   m\lambda=2\mu d\cos\theta
\end{equation}
where $m$ is an integer known as the order of interference. Then, for using a FP, additional mid-band filters, called order-sorter filters (OSF), are required for selecting only one order. The FWHM of the order sorter drives the maximum order $m$ that can be observed for a given wavelength. For OSIRIS red TF, the FWHM of the OSFs were defined so that the central 8 arcmin FOV was free from other orders for TF 
FWHM $\geq12$\AA. In other words, the minimum red TF FWHM of 12\AA\ is driven by the OSFs. However, this applies to a general case. There are specific cases, depending on the position of the wavelength of the line with respect to the wavelength range of the OSF, where it is possible to observe at even 6\AA\ FWHM, without other orders entering the 8 arcmin FOV.

The wavelength spacing between consecutive orders, known as the interÐorder spacing or free spectral range (FSR), is 

\begin{equation}
\Delta\lambda=\frac{\lambda}{m}
\end{equation}
which is obtained from equation (2) by setting consecutive integer values of $m$. Each passband has a bandwidth ($\delta\lambda$), full width at half peak transmission (FWHM), given by

\begin{equation}
\delta\lambda=\frac{\lambda(1-R)}{m\pi R^{1/2}}
\end{equation}
derived from equation (1). The ratio of interÐorder spacing to bandwidth is called the finesse, $N$,

\begin{equation}
N=\frac{\Delta\lambda}{\delta\lambda}
\end{equation}
For an ideal FP, it is given by

\begin{equation}
N=\frac{\Delta\lambda}{\delta\lambda}=\frac{\pi R^{1/2}}{1-R}
\end{equation}
whose wavelength dependence is defined by that of $R$. In most FPs the finesse can be considered approximately constant. Thus, we can see that the resolving power of a FP is equal to the product of the order and the finesse

\begin{equation}
\frac{\lambda}{\delta\lambda}=mN
\end{equation}

In other words, the FWHM depends on the order and the wavelength. OSIRIS OSFs are designed so that a minimum FWHM of 1.2 nm can be observed, free of other orders, within a FOV of 8 arcmin diameter, centred at the optical centre of the instrument, that is near the centre of the OSIRIS FOV.

The Jacquinot spot (sometimes termed the monochromatic FOV, although it is not properly monochromatic) is defined to be the region over which the change in wavelength does not exceed by $\sqrt{2}$ times the FWHM. The semi angle $\varphi$ subtended by this region is, then, from equations (2), (3) and (5),

\begin{equation}
\varphi^2\approx 2\frac{\sqrt{2}}{mN}
\end{equation}

A tunable filter is a low resolution FP where the distance $d$ is of the order of microns instead of hundreds of microns. Then, it results from equation (2) that the wavelength of a TF is tuned, for the same order, by changing the gap by nanometers, while the order (and hence the FWHM) can be changed by varying the plate spacing by microns. As a consequence, the order of interference is much lower than that of typical FPs, and since the Jacquinot spot (eqn. 8) depends mainly on $m$ for a given TF, the monochromatic FOV is larger, behaving more like a narrow-band filter for imaging emission or absorption lines, than a high resolution FP. For example, $\varphi\sim 2'$ for the OSIRIS red TF.

\subsection{Theoretical wavelength dependence across the FOV}

For an instrument such as OSIRIS, consisting of a collimator plus a camera, with the TFs located in the collimated beam, light coming from targets at increasing distances from the OSIRIS optical centre on the GTC focal plane, reach the TF at increasingly incident angles, with symmetry with respect to the optical centre. The light rays exit the TF, and reach the OSIRIS camera, with the same angle of incidence. Then, according to equation (2), there is a progressively increasing shift to the blue of the wavelength transmitted in the instrument focal plane, as the distance $r$ to the optical centre increases. However, since the beams coming from the same point on the telescope focal plane are themselves parallel, the FWHM is very nearly the same \citep{b1}.
For the case of normal incidence ($\theta = 0$), $d$ is equal to half the product of the wavelength $\lambda_0$ times the order of interference (eqn. 2). This $\lambda_0$ is the tuned wavelength, also called the Òcentral wavelengthÓ, since the TFs tuning is calibrated very near the optical centre of the instrument, where the incidence is normal \citep{b9}.

The incident angle $\theta$ at which the rays reach the TFs, and then the OSIRIS camera, depends only on the ratio between the telescope $f_{\rm GTC}$ and the instrument collimator mirror $f_{\rm Coll}$ focal lengths, and the distance $r$ to the optical centre of the instrument:

\begin{equation}
\left(\frac{\theta}{\rm deg}\right)=\frac{f_{\rm GTC}}{f_{\rm Coll}}r=2.2818\left(\frac{r}{\rm arcmin}\right)
\end{equation}
where the measured corresponding focal lengths are $f_{\rm GTC} = 169888 \pm 2$mm  and $f_{\rm Coll} = 1240.90\pm 0.05$mm. Wavelength or temperature variations can be neglected, since OSIRIS collimator is a mirror made of Zerodur, and the camera has demonstrated to be highly achromatic during commissioning \citep{b6}.

From equation (9) it follows that the light coming from a target at the limit of the OSIRIS TFs FOV (4 arcmin) reach the TFs with an angle of $9.13\degr$. It is worth noting that although larger distances are possible, the order sorter filters are designed to guarantee that other orders of interference do not leak only within a FOV of 8 arcmin diameter around the optical centre.

From equations (2) and (9), the following theoretical wavelength dependence across the OSIRIS TFs FOV can be derived:

\begin{align}
\lambda(r)&=\lambda_0\left[1-\frac{1}{2}\left(\frac{f_{\rm GTC}}{f_{\rm Coll}}r\right)^2\right]\notag \\
&=\lambda_0\left[1-7.9300\times 10^{-4}\left(\frac{r}{\rm arcmin}\right)^2\right]
\end{align}
where $r$ can be obtained from the OSIRIS plate scale $s$, of 0.127 arcsec/pixel, as measured during commissioning. This equation is an approximation that comes from the following full expression:

\begin{align}
\lambda(r)&=\lambda_0\left[\sqrt{1+\left(\frac{f_{\rm GTC}}{f_{\rm Coll}}r\right)^2}\right]^{-1}\notag \\
&=\lambda_0\left[\sqrt{1+1.5860\times 10^{-3}\left(\frac{r}{\rm arcmin}\right)^2}\right]^{-1}
\end{align}

The maximum discrepancy between equations (10) and (11) is of $\sim 2.2$\AA\ (i.e. of the order of the tuning accuracy, that is of $1-2$ \AA), at the maximum distance of 4 arcmin, and at the maximum wavelength of 9300\AA\ achievable with the red OSIRIS TF. For example, is of $\sim 1.6$\AA\ at the maximum distance of 4 arcmin, and at a wavelength of 7000\AA. The above equations could be expressed as well as a function of the OSIRIS camera focal length, since  $f_{\rm Camera}=f_{\rm Coll}/(sf_{\rm GTC})$.

\section{The observed anomalous phase effect of the OSIRIS red TF}

After starting the OSIRIS red TF scientific operations and before the end of the OSIRIS commissioning, \citet{b15} reported, for the first time, a wavelength variation across the FOV of the TF of the Magellan-Baade 6.5m telescope, depending both on the TF gap ($Z$ in units of a modified TF controller CS-100 \citep{b7}, that is equivalent to  $d$ in equation (2), in physical distance), and wavelength. As a direct consequence of the \citet{b15} results, the wavelength vs. position relation in the final TF image would not be the one derived from the incident angle over the TF, that is driven by the focal distance of the telescope and that of the instrument collimator (eqns. 10 or 11), but a general function of gap and wavelength instead. This was confirmed for OSIRIS at GTC by \citet{b13} in a private communication, who reported to the OSIRIS instrument scientific team discrepancies from the theoretical value obtained from equation (10) up to several tens of \AA, at a wavelength of $\sim 9100$\AA, while such a large departure was not evident at shorter wavelengths, of about $\sim 6500$\AA.

The dependencies with gap and wavelength detected by \citet{b15}, for the Magellan-Baade TF, clearly indicate that this phenomenon is related to the TF itself and to their inner reflective coatings. Then, it can be concluded that the origin of this Òanomalous phase effectÓ is that TFs do not behave as predicted by theoretical simple models. It is important to note that this effect cannot be physically interpreted as a change in the focal length of OSIRIS camera due to the TF, even as a function of wavelength, because the camera focal length is fixed, and with a weak wavelength dependence, as measured in the laboratory. Then, the expression providing the wavelength variation across the FOV cannot be properly described by simply changing the constants in equations (10) or (11), even as a function of wavelength.

Some OSIRIS observing projects do not require an exact knowledge of the wavelength versus position on the FOV. For example, programs devoted to the observation of objects within 1 to 2 arcmin from the optical centre, where the wavelength variation is minimum, or those whose aim is only detecting emission lines in groups, clusters or neighbouring objects. Other projects, however, do require knowing $\lambda$ from the pair or measured values $(\lambda_0, r)$.

For compensating the wavelength dependency across the FOV, it is customary to sample this variation by obtaining series of images for different values of $\lambda_0$, so that the variation in $\lambda_0$ scans the wavelength variation within the desired field around the wavelength of interest. The sampling interval ranges usually between $1/2 - 1$ times the TF FWHM. Since the narrowest FWHMs achievable with the red TF are of 12 \AA, defined by the FWHM of the OSFs, the desired accuracy following the Nyquist-Shannon theorem should be of half the sampling interval or $\sim 3-6$ \AA\ in the most demanding case (i.e. the narrowest red TF FWHM). Moreover, this accuracy is of the order of the wavelength change induced by even small dithering patterns, especially near the edges of the OSIRIS TF field. Then, in those cases where dithering is used, this accuracy will be enough. Also, for projects where an extended target is placed completely outside the optical centre, and the central wavelength $\lambda_0$ is tuned so that the centre of the target is at the desired wavelength, an accuracy of about half the red TF FWHM, or $\sim 6$ \AA, again in the most demanding case, suffices. For example, for redshift determination in blind surveys, or for estimating velocity fields in galaxy clusters, since $\Delta z/(1+z)=\Delta\lambda/\lambda=1/{\bf R}$, with ${\bf R}$ being the spectral resolution of the TF. 

\section{General equations for a Fabry-Perot}

The above derivations assume that the inner coatings are infinitely thin and with the same behaviour at all wavelengths. However, real coatings have finite thickness. Moreover, multilayer coatings, of wide wavelength range, could be rather thick, of the order of the cavity for tunable filters. Then, a very general expression for the interference condition in such a cavity is,

\begin{equation}
m\lambda=\tau_a(\lambda,\theta)+\tau_c(\lambda,\theta)=2\mu d\cos\theta+\tau_c(\lambda,\theta)
\end{equation}
where $\tau_a$ is the optical path of the air gap and $\tau_c$ is the net optical path difference (OPD) of both reflective coatings, potentially a function of wavelength and incidence angle, whose expression depends on the etalon design and its operation regime.

Equation (12) also appears in the literature in terms of the reflective phase (but usually without considering the incidence multiplicative factor, $\cos\theta$), namely:

\begin{equation*}
m\lambda=2\mu d\cos\theta-\lambda\frac{\epsilon(\lambda)}{\pi}\cos\theta
\end{equation*}
Both representations are equivalent, but here we will prefer expressing it in terms of the effective OPD of the coating, $\mu_c(\lambda)d_c$, for a symmetric comparison with the OPD in the gap Ðair- medium, $\mu d$. Later on, we will use the phase representation when deriving the profile shape of the interference peaks, when the dispersion of the reflective phase is taken into account. So, equation (12) can also be expressed as:

\begin{equation*}
m\lambda=2\mu d\cos\theta\left[1+\frac{\mu_c(\lambda)}{\mu}\frac{d_c}{d}\right]
\end{equation*}
Where it can easily be noticed that there are two regimes: the Fabry-Perot mode (high orders with large gaps, ($d_c/d\ll 1$); and the Tunable Filter regime, used at lower orders, with narrower gaps comparable to the effective coating thickness ($d_c/d\approx 1$), specially for reflective coatings for wide wavelength ranges.

\subsection{Role of coatings}

The net OPD of equation (12) depend on the coating design, specific for each FP. In this section we will briefly review different possibilities in order of complexity for the term $\tau_c(\lambda,\theta)$ in equation (12).

\subsubsection{Wide-gap Fabry-Perot approximation}

In this case $\tau_c(\lambda,\theta)=0$. Although the phase shift (per reflection), $\epsilon(\lambda)$ is ignored here, this is the most commonly used expression for etalons. Typically, shifting the actual order by one unit, it is valid when the gap separation is large compared to the coating thickness (high resolution FPs) or, regardless of the gap regime, when the reflective coatings are non dispersive in the wavelength range of interest.

\subsubsection{Ideal etalon with constant (non-dispersive) reflective phase}

Then the OPD of the coatings is a constant times the phase change $\epsilon$. An $\epsilon=180\degr$ phase change per reflection (with the constant $= 1$ for the two reflections on a FP) should in principle be taken into account for uncoated ($\mu_1/\mu_2<1$ at lower-than Brewster incidences, where the sub indices represent each reflective coating) or semi-metallic etalons. Most dielectric mirrors are designed as alternating paired quarter-wavelength thick coatings, and also produce a $\pi$ phase shift per reflection around the design wavelength. Whenever the mirror dispersion can be neglected, this is a more correct expression than the previous one, also accounting for etalons used in reflection or with a non-integer reflective phase.

\subsubsection{FP with a dispersive reflective-phase }

A dispersive reflective-phase does not change with the incidence angle, and it can be written,

\begin{equation}
\tau_c(\lambda,\theta)=2\mu_c(\lambda)d_c=\frac{\lambda\epsilon(\lambda)}{\pi}
\end{equation}
This expression is how low-order etalons (like tunable filters) are usually treated (e.g. \citealt{b1}), but breaks down in wide-field applications, where the range of incidence angles is relevant ($\Delta\theta > 1\degr$), like in the OSIRIS TF on the GTC, the MMTF of Magellan \citep{b15}, and other astronomical instruments. The expression converges to the wide-gap FP equation when the coating thickness is negligible relative to the air-gap  $d_c\ll d$.

\subsubsection{Simple TF model}

The coating is considered as an effective extra cavity,

\begin{equation}
\tau_c(\lambda,\theta)=2\mu_c d_c\cos\theta_c=2d_c(\mu^2_c-\sin^2\theta)^{1/2}
\end{equation}
As further shown, this is at least the level of characterization needed for the OSIRIS TF and similar etalons. Again, this expression converges to that in section 2.1 when $d_c\ll d$.

\subsubsection{Low-loss dielectric mirrors}

They are made of alternating layers (typically $\lambda/4$-thick) of materials with higher ($\mu_h$) and lower ($\mu_l$) refractive indices. Then two (effective) terms are required when the dispersions of both media show very different wavelength functionalities, when there are close resonant wavelengths or, after interpreting each term as a single effective medium, when the two mirrors have different coatings or multilayer structures.

\begin{equation}
\tau_c(\lambda,\theta)=2d_l(\mu^2_l-\sin^2\theta)^{1/2}+2d_h(\mu^2_h-\sin^2\theta)^{1/2}
\end{equation}

\subsubsection{More complex behaved coatings}

More elaborated characterizations require considering the detailed prescription of the multilayer coatings and/or potentially relevant effects like unparallel plates, and well-mixed, photonic or tapered coatings, among others. Some of these effects are actually present in the OSIRIS TFs and likely in many other instruments using wide wavelength, wide-field etalons at micron spacings.

\subsection{Reflective phase and the Free Spectral Range}

The reflective phase $\tau_c(\lambda,\theta)$, actually its dispersion (or derivative), also determines directly the chromatic behaviour of the FSR. Taking the difference of equation (12) at the wavelengths of two consecutive orders with a fixed gap (and incidence), we have

\begin{equation*}
m\lambda_{m}-(m+1)\lambda_{m+1}=\tau_c(\lambda_m,\theta)-\tau_c(\lambda_{m+1},\theta)\equiv\Delta\tau_c
\end{equation*}
then

\begin{equation*}
\Delta\lambda\equiv\lambda_m-\lambda_{m+1}=(\lambda_{m+1}+\Delta\tau_c)/m
\end{equation*}
or

\begin{equation}
\Delta\lambda=\frac{\lambda_m+\Delta\tau_c}{m+1}
\end{equation}
Therefore, whenever $\Delta\tau_c\ll \lambda$, either because the air-gap is much larger than the coating 
($\mu_c(\lambda)d_c\ll \mu(\lambda)d$),  the FSR can then be expressed in the more known simple form of equation (3):

\begin{equation*}
\Delta\lambda\equiv\lambda_m-\lambda_{m+1}=\lambda_{m+1}/m=\lambda_m/(m+1)
\end{equation*}
for $d_c/d\ll 1$ or $\tau_c\neq f(\lambda)$, i.e. $\epsilon\propto 1/\lambda$.
The first term of equation (16) corresponds to the large-spacing case, that is traditionally used in reference books, but that is clearly a poor approximation in a TF regime with a dispersive reflecting coating. Also, the traditional approximation still holds with a reflective phase that is constant in $\lambda$ ($\delta\epsilon/\delta\lambda=0$). However, in this general formulation, the dispersion of the dielectric coating adds a $\lambda$-dependent correction to the simple standard approximation. Furthermore, in this additional term, 
$\Delta\tau_c=\tau_c(\lambda_m)-\tau_c(\lambda_{m+1})$
is scaled by the incidence factor $\cos\theta$, a feature that allows the determination of the controller zero-point ($d$ in equation 12) when the etalon is characterized at a pair of known incidences (not just normal incidence as is typically done).

\subsection{Reflective-coating effects on the spectral resolution}

The resolving power ({\bf R}) of the etalon is now limited by the effective finesse of the etalon:

\begin{equation}
{\bf R}(\lambda,m)=\frac{\lambda}{\delta\lambda}=mN_{\rm eff}(\lambda)
\end{equation}
If we take into account the reflective phase of equation 12 in the standard derivation of {\bf R}, the general expression for the resolving power explicitly depends on the $\lambda$-derivative of the reflective phase (i.e., the actual chromatic dispersion of the coating):

\begin{equation}
\frac{\lambda}{\delta\lambda}=N_{\rm eff}(\lambda)\left(m-2\frac{d\tau_c}{d\lambda}\cos\theta\right)
\end{equation}
Just like for the FSR case, the expression for the FWHM includes an additive term that goes as the coating dispersion scaled by the incidence factor.

\subsection{Reflective-coating effect at different incidences (phase effect)}

When we tune our etalon for a wavelength $\lambda_0$ in the central field ($\theta=\theta_0$ or $\theta=0$ for normal incidence or a non-inclined etalon), we apply the constructive interference condition and are left with a discrete set of orders $m$ to choose from, based on the most desired resolving power. The proper $d - \lambda$ relation that considers the reflective phase to set the corresponding gap thickness is the given by equation (12), rewritten as

\begin{equation}
d\equiv d(\lambda_0,m)=\lambda_0\frac{m}{2\mu}-\frac{\tau_c(\lambda_0)}{\mu}
\end{equation}

The phase delay of the reflective coatings makes then the actual $\mu$-gap of a TF to be narrower than predicted by the traditional relation (1st-term only: $d=(m/2\mu)\lambda_0$) that ignores its effects, applicable to wide-gap FPs. This effect was already detected, albeit unexplained, during the factory validation of OSIRIS blue TF (it was less conspicuous for the red TF). 

Once the central wavelength $\lambda_0$ has been tuned, the remaining field angles within the field-of-view  of our instrument, feed the etalon at different incidences. For a given incident field angle $\theta$, the wavelength $\lambda(\theta)$ that undergoes constructive interference (just a single one in the same $m$) is the one that satisfies equation (12). The general expression applicable to both, FP and TF regimes, for the field-dependent phase difference, $\lambda(\theta)-\lambda_0$, is then directly derived

\begin{equation}
m(\lambda(\theta)-\lambda_0)=2\mu d(\cos\theta-1)+2\tau_c(\lambda(\theta))\cos\theta-2\tau_c(\lambda_0)
\end{equation}
which, after substitution of the tuned gap from equation (19) ($2\mu d=m\lambda_0-2\tau_c(\lambda_0))$, becomes

\begin{equation}
\lambda(\theta)=\lambda_0\cos\theta+\frac{2}{m}(\tau_c(\lambda(\theta))-\tau_c(\lambda_0))\cos\theta
\end{equation}
The second term on the right can be seen as the coating dispersion correction to the field-phase that, in the TF regime ($m\sim 10, \tau_c\sim 1, d_c/d\sim 1$) can be comparable or even larger than the $\lambda_0$-term, the only one traditionally considered in the literature, and applicable to the FP regime ($m\gg 1$ or $d_c/d\ll 1$). Then again, as was the case for the FSR, equation (21) shows how the single $\lambda_0$-term approximation is also valid in case of a finite but non-dispersive reflective phase ($\tau_c(\lambda)=\tau_c(\lambda_0)$), even in the TF regime.

Depending of the wavelength dependence of the coating dispersion, the reflective phase correction can render equation (21) to be transcendent in $\lambda(\theta)$. In the next section we derive the proper analytic expression $\tau_c(\lambda)$  that should be used to resolve equation (21). 
In the meantime, we can use the approximation

\begin{equation}
\Delta\tau_c=\tau_c(\lambda(\theta))-\tau_c(\lambda_0)\approx \Delta\lambda(\delta\tau_c/\delta\lambda)_0=(\lambda(\theta)-\lambda_0)\tau'_c(\lambda_0),
\end{equation}
valid when the $\lambda$-shift over the field is much smaller than the central wavelength, to get an un-transcendental expression for the field-phase (approximation to equation 21) in terms of the $\lambda$-derivative of $\tau_c$, evaluated at the central wavelength as tuned:

\begin{equation}
\lambda(\theta)-\lambda_0\approx \lambda_0\frac{\cos\theta -1}{1+(2/m)\tau'_c(\lambda_0)\cos\theta}
\end{equation}

In the following section we derive the chromatic OPD, $\tau_c(\lambda)$ of the OSIRIS red tunable filter, and further in this work we derive from them the proper analytical function $\lambda(\theta)$ that applies. Since the reflective coatings of the OSIRIS TF are found to be moderately dispersive ($d\tau_c/d\lambda\neq0$), particularly at longer wavelengths, the measured $\lambda(\theta)$  at the telescope deviates from the traditional standard formula, just as predicted by equations (21) or (23) (further section).

\section{Measuring the effective phase and its dispersion}

\subsection{Basic equations}

From the standard laboratory characterization of an etalon, namely a mapping of the interference peaks using a scanning monochromator, the simple but general formulation above allows directly deriving the effective OPD,$\tau_c(\lambda)=2\mu_c d_c$, of the reflective coating with very high accuracy, without the knowledge of the detailed prescription of the multilayer coating.

Figure 1 shows the mapping of the transmission peaks of the OSIRIS red TF, as a function of the digital unit, $Z$, of the modified CS-100 piezoelectric controller of the gap thickness ($d_a \propto 1/Z$) as measured in the lab at normal incidence \citep{b9}. At a constant gap (fixed $Z$), these are the wavelengths of consecutive orders that, for a non-dispersive reflective phase, should perfectly collapse in a single narrow curve on an 
$m\lambda$ vs. $Z$ diagram, provided that the initialization of each order sequence is properly identified. For a linear controller, like the CS-100, the curve should be a line, but indeed the rms dispersion from the curve does not reduce under a quadratic or higher order assumption for the air-gap relation to $Z$.

The presence of a chromatic reflective phase and the need to account for it, is shown in Fig. 2, where very large systematic wavelength departures, up to a few tens of nanometers, results from a non zero, non linear $\tau_c$ term in equation (12).
 
\begin{figure}
\includegraphics[width=0.35\textwidth,angle=270]{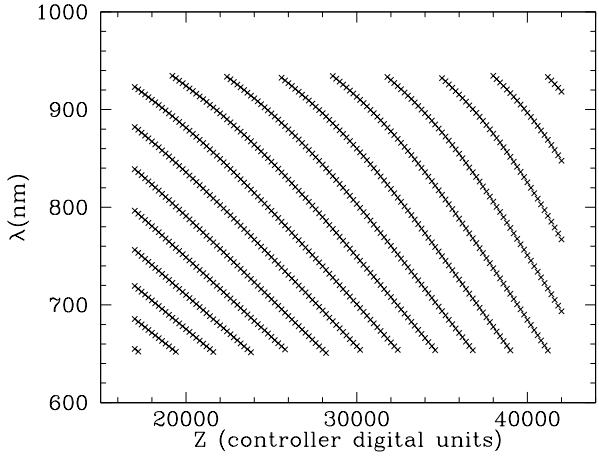}
\caption{Transmission peaks of the OSIRIS red TF as a function of controller digital units as measured in the laboratory at normal incidence. Note the non-linear shape of the curves due to the anomalous phase effect.}
\label{schema2}
\end{figure}

\begin{figure}
\includegraphics[width=0.35\textwidth,angle=270]{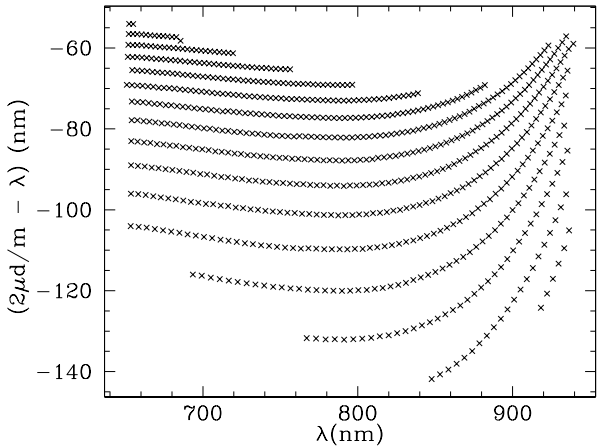}
\caption{Residuals of equation 12 for normal incidence. Every curve corresponds to a different value $m$, increasing from bottom to top. The offset of the different curves in the vertical axis show the presence of a phase effect due to a chromatic dispersion caused by the internal coatings of the etalon, while deviations from a straight line vs. wavelength, mainly beyond $\sim 800$nm, show the presence of an additional non-linear relation between air-gap and $Z$.}
\end{figure}

We can now use equation 19 to directly derive the effective optical path difference of the coating, 
$\tau_c(\lambda)=\mu_c(\lambda)d_c$, from the direct measurement of  the  $d(\lambda)=d_0+bZ(\lambda)$ relation. This is usually done in the lab, feeding the etalon with a collimated white-light beam and scanning, with the aid of a monochromator, the $\lambda$-location of the consecutive interference peaks for a set of $Z$ settings of the etalon controller, i.e., the so called Ò$Z(\lambda)$Ó calibration of the etalon, at a known incidence (typically normal incidence, $\cos\theta\approx 1$):

\begin{equation}
\tau_c(\lambda)\equiv\mu_c(\lambda)d_c=\lambda-2\frac{d}{m}
\end{equation}
or

\begin{equation}
\tau_c(\lambda)=\lambda-2\frac{d_0+bZ}{m}
\end{equation}

Equation (24) is useful when the controller $d(Z)$-relation is already known, while equation (25) allows us to derive at once the effective dispersion of the etalon reflective coating and the controller slope $b$. The zero point of the controller calibration, $d_0$, remains yet to be determined, since equation (25) couples it to any potential constant term in the coating refractive index $\mu_c(\lambda)$ (further we will see how the same empiric  $Z(\lambda)$ calibration, or better yet the $\lambda(\theta)$ phase calibration with field angle, also yield an extremely accurate value for $d_0$ though). For the moment, lets continue our discussion with equation (25), or once the controller calibration has been settled.

As can be seen, the empiric measurement (25) gives the coating OPD and its dispersion, directly in terms of the refractive index of the medium in the etalon (typically air).  So an effective mono-layer coating with the such derived $\mu_c(\lambda)d_c$, and with a dielectric (or imaginary part of the refractive index) that matches the -also measurable- reflectivity of the etalonÕs coating, gives a perfectly equivalent performance.

\subsection{Normal incidence fits}

At normal incidence, equation (25) can then be used to obtain both the gap-$Z$ relation and the coating OPD. Note that this relation depends on order $m$, which is unknown and must be guessed until all the curves in Fig. 2 collapse in one. In Fig 3. we show the resulting OPD that best fit into equation (25). 

\begin{figure}
\includegraphics[width=0.35\textwidth,angle=270]{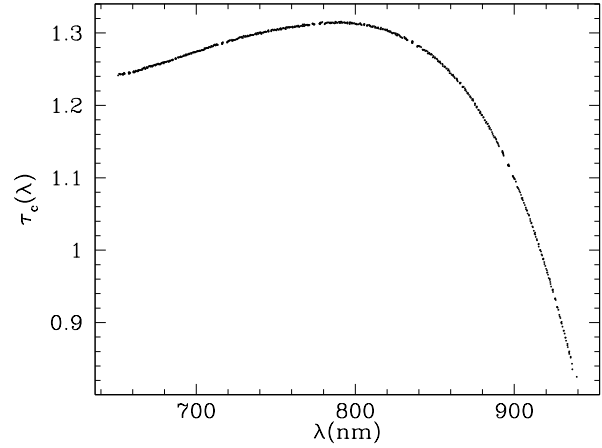}
\caption{OSIRIS red TF coating OPD obtained by fitting equation (25) to normal incidence lab data.}
\end{figure}

To completely determine the TF performance the effective finesse $N_{\rm eff}$ must be known. Once we know $\tau_c(\lambda)$, $N_{\rm eff}$ is obtained from equation  (18) through the measured FHWM, $\delta\lambda$. 
Fig. 4 shows the resulting finesse. 

\begin{figure}
\includegraphics[width=0.35\textwidth,angle=270]{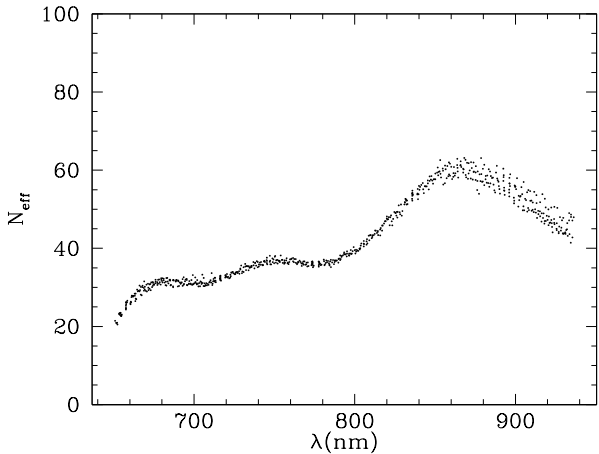}
\caption{Effective finesse of the red TF.}
\end{figure}

In summary, at normal incidence, laboratory data can be fitted by a single-coating model (equation 14) and characterize the TF performance. The obtained $\tau_c(\lambda)$ and $N_{\rm eff}(\lambda)$, can then be used to tune the central wavelength and FWHM with accuracies of $\sim 1$\AA\ and $\sim 0.25$\AA, respectively.

A full TF characterization, including the phase effect, requires studying the case for non-normal incidence. However, since now a dependency on the incident angle $\theta$ at pupil appears, this model should be fitted by illuminating the TF using the instrument optics, as discussed in the following section, combined with laboratory data. In order to do this correctly, we must take into account that the etalon is sensible to temperature changes, that is, the gap-$Z$ relation depends on temperature. In the present work, when merging lab and telescope data to try a model fit, this effect have been taken into account and corrected.

\section{Calibration techniques for non-normal incidence}

\subsection{ICM rings}

The GTC is equipped with several spectral lamps located in the Instrument Control Module (ICM), which is designed to illuminate the telescope focal plane as an $f/17$. An emission line produced by a spectral lamp can be seen as a central bright spot when the TF is tuned to the wavelength of the line (this is the system used for TF wavelength calibration, see \citep{b9}) or as a ring with centre the optical axis of the instrument, and at a certain radius $r$ on the detector, when the TF is tuned to redder wavelengths (Fig. 5). These rings could be used for calibrating the wavelength dependence across the OSIRIS TFs FOV by fitting centre and radius.

\begin{figure}
\includegraphics[width=0.45\textwidth]{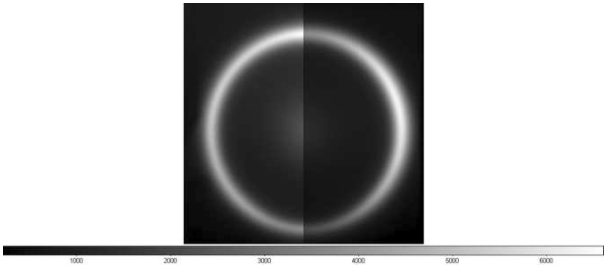}
\caption{Image of a Ne spectral lamp illuminating the red TF at a central wavelength tuned to the red of the line. A ring forms due to the central wavelength variation across the OSIRIS FOV. Both CCD detectors are shown without including the gap between them. }
\end{figure}

The main advantages of this procedure are that

\begin{itemize}
\item The lines are intense and single lines can be selected in advance
\item It can be used in day conditions, thus saving observing time
\item It can be used to reliably determine and monitor the optical centre and the tuning stability
\end{itemize}

The main disadvantages are:

\begin{itemize}
\item That the rings are not uniform because the GTC focal plane is not uniformly illuminated by the ICM. Then fitting rings could be sometimes difficult 
\item This non-uniformity depends on the lamp used, since different lamps are located in different positions surrounding the focal plane
\item At low $m$, the rings are quite thick and fitting the radius becomes more difficult 
\item At longer wavelengths the fringing on the rings makes it difficult to fit rings
\item There are not many single and strong emission lines that can be observed for the same fixed $\lambda_0$, and certainly not everywhere in the spectral domain
\item Inserting and removing the ICM mirror is a time consuming operation
\end{itemize}

\subsection{Sky rings}

In the red part of the optical sky spectra there are many emission lines that occur at quite stable wavelengths, and they could be used for calibration. An example is shown in Fig. 6. The main advantages are:

\begin{itemize}
\item Their optical path is the same than that for astronomical objects
\item They could be obtained from the same scientific data
\item If not, changing the TF tuning and the OSF is faster than inserting and removing the ICM mirror
\end{itemize}

The main disadvantages are:
\begin{itemize}
\item They must be observed during the night (unless suitable lines appear in the scientific data)
\item The lines are faint, and most of them multiple (specially considering the low spectral resolution of the TFs). Then suitable lines are scarce and difficult to fit
\item Their strengths varies with time and airmass
\item They are not present in the whole spectral domain
\end{itemize}

Given these disadvantages, sky rings can be mainly used for on-line quick monitoring the spectral stability of the TFs, not for a general calibration of the wavelength dependence of OSIRIS TFs over their FOV.

\begin{figure}
\includegraphics[width=0.45\textwidth]{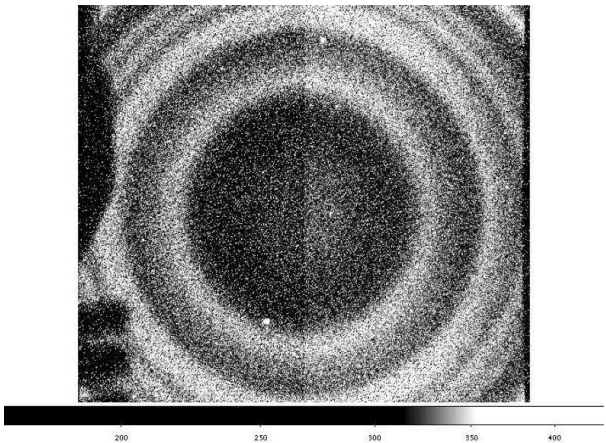}
\caption{Sky ring at 894.35 nm with the OSF 878/59 with FWHM 1.21nm, tuning the red TF at 898.2nm for obtaining a ring radius of $\sim 1100$ pixels (unbinned). The exposure time was of 120 s.}
\end{figure}

\subsection{Astronomical targets}

Emission lines from astronomical objects could be used for calibration. In this case rings do not appear since it is not the whole focal plane that is illuminated. Then, either the targets are widespread over the FOV (as for galaxy clusters) or they are placed at different suitable redshifts (as, for example, for bright QSOs) and their position varied across the FOV. Please note that since we are mainly interested in keeping the central wavelength fixed, the procedure of observing the same target in the same emission line at different r and varying $\lambda_0$, would not necessarily provide the full desired calibration. This essentially discards using galactic targets. 

The main advantages are that, known astronomical targets

\begin{itemize}
\item Illuminate the GTC focal plane as unknown targets do
\item Targets such as clusters or QSOs are quite abundant and can be chosen for calibrating specific spectral regions
\end{itemize}

The main disadvantages are:

\begin{itemize}
\item They must be observed during the night 
\item This procedure represents a significant investment of observing time
\end{itemize}

Then, calibrated astronomical targets will be mainly used for checking the general calibration of the wavelength dependence of OSIRIS TFs over their FOV derived.

\subsection{Data used for non-normal incidence calibration}

The data from the present analysis consist on:
\begin{itemize}
\item ICM rings from spectral lamps, obtained both from the same emission line varying $\lambda_0$, and for a fixed $\lambda_0$ observing different spectral lines, courtesy of A. Cabrera-Lavers (Table 1).
\item Emission line galaxies with public spectroscopy from a cluster observed with the red TF for the GaLAxy Cluster Evolution (GLACE) project \citep{b13}, that allowed both observing nearly the same wavelength for targets at different $r$ and different $\lambda_0$; and observing at the same $\lambda_0$ the same line at different $r$ and different wavelengths, due to the Doppler shift of the different targets.
\end{itemize}

The ICM rings were fitted by eye mainly over CCD2 (the detector at the right in Fig. 5) using DS9 and circular regions. These fittings allow obtaining an accuracy of $\pm 4$ pixels (binned $2\times 2$) both in centre and radius. These errors are equivalent to less than 1 \AA\ in the worst case, i.e.: smaller than the tuning accuracy. Fitting CCD1 rings were used as control to check consistency of the results. No difference was detected in centre or radius when using CCD1.

\begin{table*}
\centering
\begin{minipage}{140mm}
\caption{GTC ICM data observed through OSIRIS}
\begin{tabular}{lccccr}
\hline
Lamp\footnote{Wavelengths from the database of the National Institute of Standards and Technology http://physics.nist.gov/PhysRefData/ASD} & Line & TF $\Delta\lambda$ & RMA & $\lambda_0$&Observing data\\
&(\AA) & (\AA) & ($\degr$) & (\AA) &\\
\hline

Ne &6598.9 &12 &60,-75& 6620,6645,6660,6675&31/07/11\\
Ne &6929.5 &16 &60,-75& 6940,6960,6980,7000&18/06/11\\
  &&12,15,19& 60,-75& 6940,6960,6980,7000	&01/08/11\\
Ne&7032.4&16&60,-75& 6950,6975,7000,7020	&18/06/11\\
Hg(Ar)& 7635.1 &12& 60,-75& 7650,7665,7685,7700 &31/07/11\\
Ne & 8377.6 &12 &60,-75& 8400,8420,8440,8460 &31/07/11\\
Xe& 8819.4 &12 &60,0,-75& 8840,8860,8880,8920,8940	& 27/06/11\\
Xe& 9162.5 &12 &60,-75& 9185,9200,9215,9230 & 31/07/11\\
Ne,Hg(Ar) & 6929.5 & 12 & 60 & 7042 &21/08/11\\
  &6965.4& & & & \\
  &7032.4 & & & & \\
Xe,Hg(Ar)& 9123.0&12	&60& 9172 &21/08/11\\
&9162.5& & & & \\
\hline
\end{tabular}
\end{minipage}
\end{table*}

From the ICM data analyzed (Table 1), the following conclusions can be readily obtained:

\begin{enumerate}
\item Neither the centre nor the radius of the rings change with rotator position, where $60\degr$ correspond to the TF looking ÒupwardsÓ. Then possible TF gravitational flexures do not affect the calibration looked for
\item The centre of the rings does not show significant wavelength or order of interference, dependences, although at longer wavelengths and lower orders the fitting uncertainty increases
\item The centre and radial dependency of the rings does not vary with time (at least within the periods of time explored)
\item The centre of the ring is at (CCD2 reference pixels): $-11\pm 1, 976\pm 1$ pix
\item Distortion could affect the results up to almost 3\AA\ at the edge of the OSIRIS TF FOV. Hence all prescriptions given here must be applied BEFORE astrometry corrections
\end{enumerate}

\section{Non-normal incidence}

\subsection{Single coating model}

Our single-coating model of the etalon, given by equations (12) and (14), also accounts for the observed change of central wavelength at non-normal incidence, i.e. with distance to the optical axis. We can express the $\lambda(\theta)$ as

\begin{equation}
\lambda(\theta)=\frac{1}{m}\left[2\mu d\cos\theta+\sqrt{\tau^2_c(\lambda)-4d^2_c\sin^2\theta}\right]
\end{equation}

It is important to note that now, contrary to the case at normal incidence, there is no coupling between independent terms in air gap, $d$, and $\tau_c$. Therefore, calibration at non-normal incidence will provide the actual zero point of the gap-$Z$ relation. Combining equation (26) at both normal incidence and non-normal incidence we have,

\begin{equation}
\lambda(\theta)-\lambda(0)=\frac{1}{m}\left[2\mu d(\cos\theta-1)+\sqrt{\tau^2_c(\lambda)-4d^2_c\sin^2\theta}-\tau_c(\lambda)\right]
\end{equation}
which is then  fitted to the ICM data.

The best fit we obtain gives a rms of $0.7$ \AA\ with a coating of $1.3 \mu$m. In Fig. 7 we represent the residuals as a function of data index. This index represents the ordering of data, first by $Z$, and then by incidence angle; in this way, data indices larger than 700 correspond to non-normal incidence.  We see that there are points (at non-normal incidence) whose residuals are high ($\sim 5$ \AA). This indicates that a single coating is not enough to explain the TF behaviour, and that a more realistic model should include a multilayer coating.

\begin{figure}
\includegraphics[width=0.38\textwidth,angle=270]{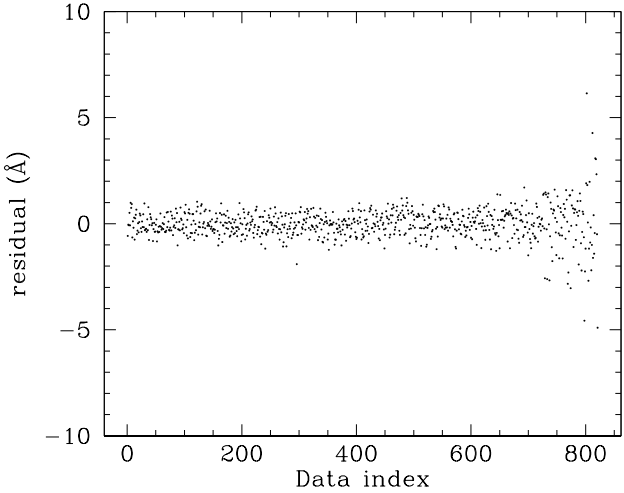}
\caption{Residuals of the fit to the single-coating model given by equation (27).}
\end{figure}

\subsection{Multiple layer coating empirical fitting}

The results of the previous section imply that, not surprinsingly, a single layer cannot properly fit the non-normal incidence case for the thick reflective coatings of the OSIRIS red TF. Then, more terms have to be added, as shown in the simplest case of two-layered coatings of equation (15).  A two-coatings model was then fitted to the data obtaining a rms of $0.5$ \AA\ with coating thickness of 1.1 and 0.05 $\mu$m. In Fig. 8 we represent the new residuals showing a clear improvement with respect to the one-coating model.

\begin{figure}
\includegraphics[width=0.38\textwidth,angle=270]{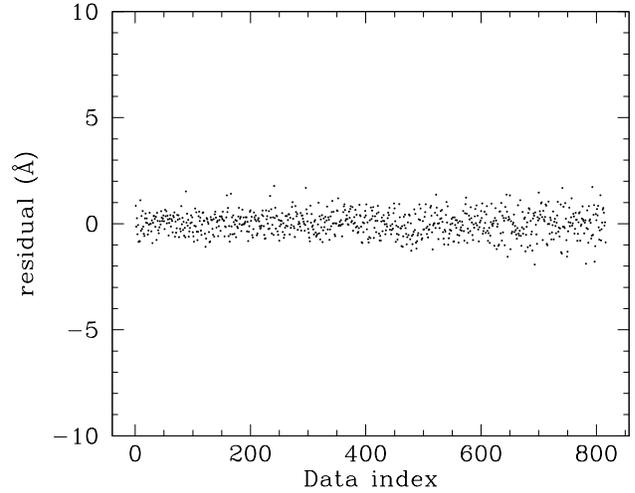}
\caption{Residuals of the fit to a two-coating model}
\end{figure}

Equation (27) is a transcendental function between wavelength and distance to the optical axis. Hence, it is difficult obtaining a $\lambda(\theta)$  solution. However, if we consider the case of low incidence angles, expression (27) can be approximated by a general (yet transcendental) form, which, at the lowest order, behaves like $\theta^2$ or, equivalently as $r^2$ according to equation (9), as in the theoretical case (equation 10), but with different coefficients, in general depending on wavelength. 

Following this procedure, from the ICM data, it results that the simple function

\begin{equation}
\lambda=\lambda_0-5.04r^2
\end{equation}
relates the distance $r$ in arcmin to the optical centre, assuming the standard scale factor (equation 9), with the wavelength $\lambda$ in \AA, fixed the central wavelength $\lambda_0$. Then, the shift to the blue from optical centre to edge of the red TF FOV does not depend on wavelength on first approximation, and is lower than the theoretical one (equation 10). 

The expressions (10) and (28) are identical for $\lambda_0\sim 6356$\AA, and equation (28) departs from the theoretical one (equation 11) in an amount that remains within the tuning error at 6563\AA, even at the edge of the red TF FOV. However, this departure increases with wavelength and radius. It reaches $\sim 6$\AA\ at 2 arcmin from the TF optical centre at 8500\AA, and up to 35\AA\ at the edge of the red TF FOV at 9300\AA. Then, the theoretical expression (11) departs, even within the monochromatic FOV (equation 8), well beyond the tuning accuracy at any wavelength beyond 6800\AA, and in excess of $\sim 6$\AA\ at 8500\AA\ and beyond.

This explains the reason why this effect remained undetected during instrument verification in the laboratory: the phase effect was measured using the Ne line 6929.5\AA\ only, and assuming an approximate calibration of 0.2\AA/bit, since the calibration wavelength versus $Z$ bits (gap in CS-100 units) was not yet available when this measurement was taken. Since in this spectral region the departures with respect to the theoretical expression are moderate, of $\sim 7$\AA\ at the edge of the 8 arcmin diameter FOV, the uncertainty in the wavelength calibration masked this departure. Finally, since the scientific operation started before finishing the red TF commissioning, tests such as those proposed by \citet{b5}, that would have revealed the problem, were not performed.

Within the monochromatic FOV, equation (28) provides very good accuracy at any wavelength. Although (28) looses accuracy as $r$ increases, depending on wavelength, the overall fit error and the maximum error at the edge of the red TF FOV are within the tuning error ($1-2$ \AA) (see Fig. 9) in most cases. Significant departures are located around 6500, 7600 and 9200\AA, where the errors are up to $\sim 4-5$\AA\ in the worst cases (Fig. 10). It is worth noting that this error and the tuning error shall be added in quadrature, since they are independent. As a conclusion, expression (28) is accurate enough for most applications, specially when observing using a dithering pattern, as discussed in Section 3.

\begin{figure}
\includegraphics[width=0.35\textwidth,angle=270]{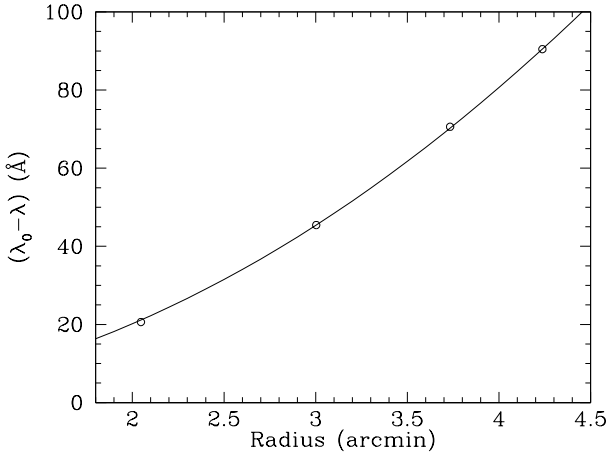}
\caption{$(\lambda_0-\lambda)$ vs. radius for the Ne line 6929 \AA. The curve is that given by equation (28). The fitted tuning error is of 0.2 \AA, the error of the overall fit is of 0.5 \AA.}
\end{figure}

\begin{figure}
\includegraphics[width=0.35\textwidth,angle=270]{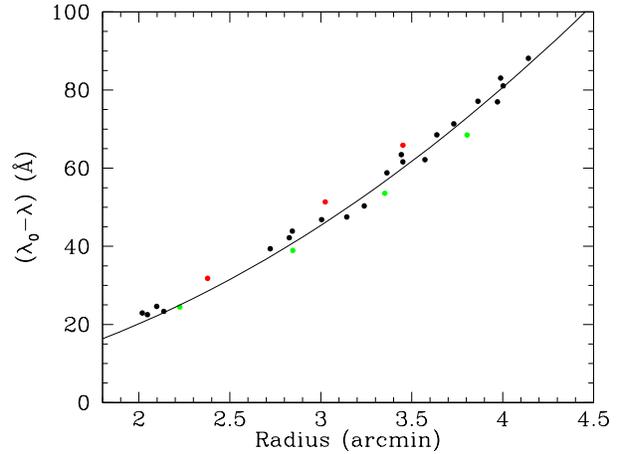}
\caption{$(\lambda_0-\lambda)$ vs. radius for all ICM data. Red points correspond to 7635\AA\ and the green ones to 9162\AA. The curve is that given by equation (28).}
\end{figure}

However, when more accuracy is required, specially at those wavelengths where the departure is higher, an additional chromatic term expressing the wavelength dependence of the coatings $a_3(\lambda)$ can be added,

\begin{equation}
\lambda=\lambda_0-5.04r^2+a_3(\lambda)r^3
\end{equation}
where
\begin{equation}
a_3(\lambda)=6.0396-1.5698\times 10^{-3}\lambda+1.0024\times 10^{-7}\lambda^2
\end{equation}
with $\lambda$ in \AA

This equation, that is non-linear, as expected, has been obtained from the different fits to the different emission lines considered (Fig. 10), using a Levenberg-Marquardt multiple polynomial fitting, and provides accuracy of the order of the tuning error $a_0$, whose value is not included in the above expressions (Fig. 11). Equation (30) can be applied iteratively for a real case where $\lambda$ is the unknown, or it can be approximated by considering $\lambda_0$ in (30) instead since the differences between $\lambda_0$ and $\lambda$ are, at most, of $\sim 80$\AA\ at the edge of the red TF FOV.

Keeping a second order polynomial (i.e. $a_3=0$) and assuming an $a_2(\lambda)$ term, yields worse results in the most extreme case ($\lambda=7635$\AA, Fig. 12), although the results are as good as with equation(29) in the remaining cases (Fig. 13).

\begin{figure}
\includegraphics[width=0.35\textwidth,angle=270]{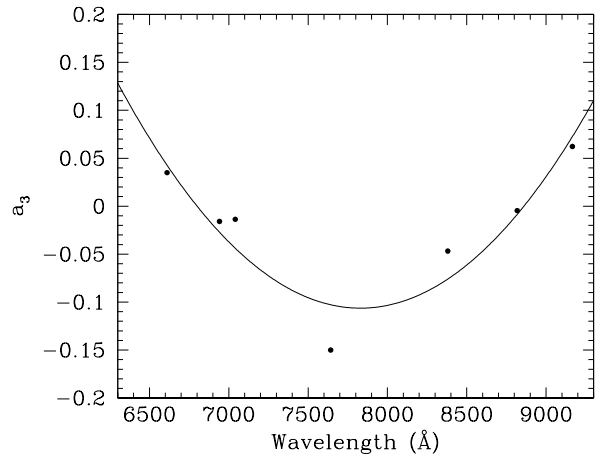}
\caption{Coefficient $a_3$ vs. wavelength. The fitting provides equation (30)}
\end{figure}

\begin{figure}
\includegraphics[width=0.35\textwidth,angle=270]{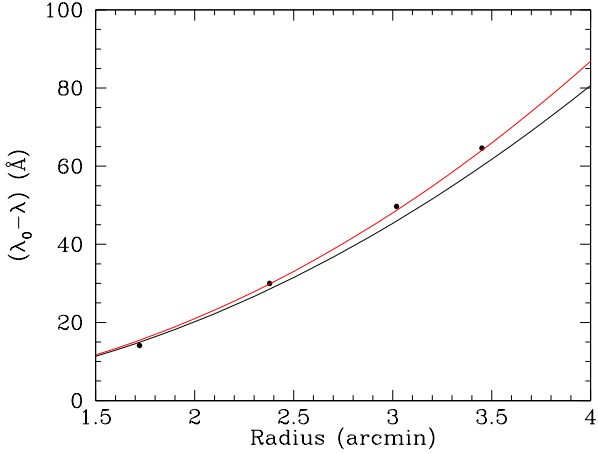}
\caption{$(\lambda_0-\lambda)$ vs. radius for the HgAr line 7635\AA. The blue curve is that given by expression (28) while the red one is that obtained using equation (29). The fitted tuning error of the red curve is of 0.5 \AA, and the error of the overall fit is of 1.1 \AA.This is the worst case analyzed.}
\end{figure}

\begin{figure}
\includegraphics[width=0.35\textwidth,angle=270]{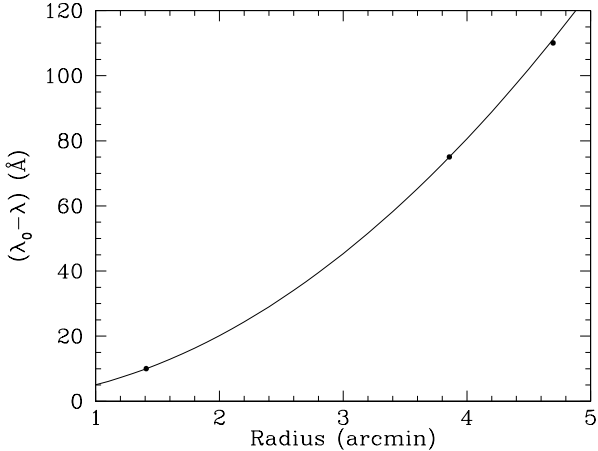}
\caption{$(\lambda_0-\lambda)$ vs. radius with $\lambda_0$ fixed at 7042\AA, for the spectral lines Ne 6929\AA, HgAr 6965\AA, and Ne 7032\AA\ using equation (28). The tuning error fitted is of 1 \AA\ and the overall fitting error is of 1.5 \AA. It improves to 0.8 \AA\ using model (30) with $\lambda_0$. However, in this case, in the range $6929-7032$ \AA, the value for $a_3$ does not change significantly}
\end{figure}

However, as already discussed in Section 3, most programs will not require such an accuracy, since if a dithering pattern is used for removing TFs ghosts and other spurious, the wavelength at which targets near the edge of the OSIRIS red TF FOV are observed, changes by more than 6\AA\ for a dithering of 10 arcsec (e.g. for removing the gap between both OSIRIS CCDs).

Finally, by definition, the radius of the Jacquinot spot is, then,

\begin{equation}
\left(\frac{\varphi}{\rm arcmin}\right)^2\approx \frac{\sqrt{2}\delta\lambda}{5.04}.
\end{equation}

To compare this last fit with the two-coatings models we calculated the difference of the residuals from both fits. Figure 14 shows this difference for four central wavelengths. This shows that the approximation given by equations (29) and (30), being easier to use, is fully compatible with our two-coatings model, which explains the behaviour of a low-order etalon.

\begin{figure}
\includegraphics[width=0.35\textwidth,angle=270]{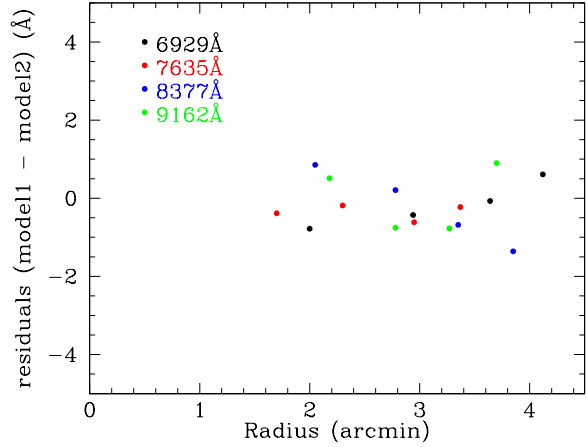}
\caption{Difference of residuals between the two fits discussed here as a function of radius along FOV.}
\end{figure}

\subsection{Analysis of target lines}

The expressions derived above were checked using H$\alpha$ data of emission line galaxies (ELG) belonging to the rich galaxy cluster Cl0024+1654 at z\,=\,0.395, within a single OSIRIS red TF pointing. The observations were performed in the framework of the GLACE survey (S\'anchez-Portal et al. 2013;  S\'anchez-Portal et al., 2014, hereafter MSP14),  aimed at targeting  several clusters at different redshifts (z\,$\sim$\,0.40, 0.65 and 0.85) to perform a thorough study on the variation of galaxy properties (star formation, AGN activity and morphology) as a function of environment and cosmic time. The open and guaranteed time observations were carried out in two observing campaigns (GTC semesters 09B and 10A). The spectral range 9047--9341\,\AA\  was covered by 50 evenly spaced scan steps ($\Delta\lambda$\,=\,6\,\AA) with a total of 5.15 hours of on-source integration time.  At each TF tune, three individual exposures with an ``L'' shaped dithering pattern of 10\,arcsec amplitude were taken (in order to allow the removal of fringes and to ease the identification of diametric ghosts). The data reduction was performed using a version of the TFRED package \citep{b14} modified for OSIRIS by us and private IDL scripts written within our team. The data reduction procedure will be thoroughly described in  MSP14, but it is outlined here for reference. The basic reduction steps includes bias subtraction and flat-field normalization followed by removal of the sky rings in each individual exposure by means of a background map created by computing the median of several dithered copies of the object-masked image. Fringing is also removed if required using the dithered images taken with the same TF tune. Then, the frames are aligned and astrometrically registered and a deep image is obtained by adding up all individual exposures of every scan step. This deep image is used to extract the sources by means of the SExtractor package \citep{b17}. 
For each detected source and scan step, the best possible combination of individual images, i.e. the best combinations of TF tune and dither position at the location of the source (in practice all the images for which the TF wavelength at the position of the source lies within a range of 6\,\AA\ of the given one) is derived and the output combined image is used to determine the flux at this specific scan step and source position by means of SExtractor. The resulting pseudo-spectra consist of 50 
tuples ($\lambda$ at source position, flux). In many cases, the H$\alpha$ and [N{\sc ii}] lines appear clearly resolved. The position of the H$\alpha$ line for each source is then determined by standard spectral fitting procedures (for instance the IRAF\footnote{IRAF is distributed by the National Optical Astronomy Observatories, which is operated by Association of Universities for Research in Astronomy, Inc., under cooperative agreement with the National Science Foundation.} \texttt{splot} task).
From the H$\alpha$ emitters detected in a single OSIRIS red TF 
pointing, 72 had spectroscopic redshifts from \citet{b11}. These are indicated with red circles in Figure 
15 where a broad-band image of this cluster is shown. From these, 6 were detected at the same mean TF 
central wavelength 9173.4 \AA\ and 6 more at 9197.4 \AA, while 7 were observed at the same wavelength 
($9160\pm 2$ \AA) but at different TF central wavelengths. Many of the spectroscopic redshifts 
of the sources are classified as ÒuncertainÓ by \citet{b11}, and then some dispersion is expected. 
However, given the relatively large amount of targets, the results should be significant at least 
when using the full dataset.

The behavior of equations (28) and (29) for the whole set of data, except 6 spurious galaxies, is 
presented in Fig. 16. Note that this plot does not represent the velocity field of the cluster, which is discussed in \citet{b18}. The dispersion observed is mainly attributable to  \citet{b11} spectroscopy, 
plus the errors associated to the dithering procedure, that increase with radius. A good fitting is 
obtained to the mean value of the dataset.

\begin{figure*}
\includegraphics[width=0.7\textwidth,angle=0]{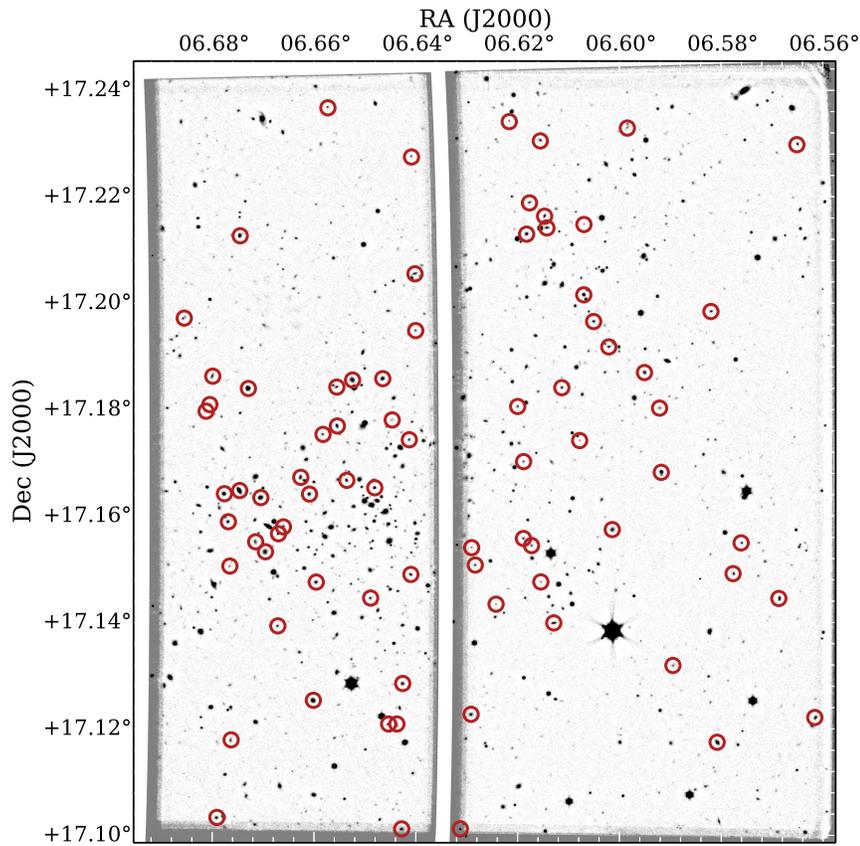}
\caption{OSIRIS red TF deep image of cluster Cl0024+1654 (z = 0.395) at the redshifted H$\alpha$ line.
Galaxies with spectroscopy from \citet{b11}, used in this work, are marked with red circles. As explained in the text and in MSP14, this deep image has been obtained combining many dithered images using a median filter. This method is very effective to remove artifacts (as compared to other combination methods) and therefore no ghost images are expected, but only real detections.}
\end{figure*}

\begin{figure*}
\includegraphics[width=0.7\textwidth,angle=270]{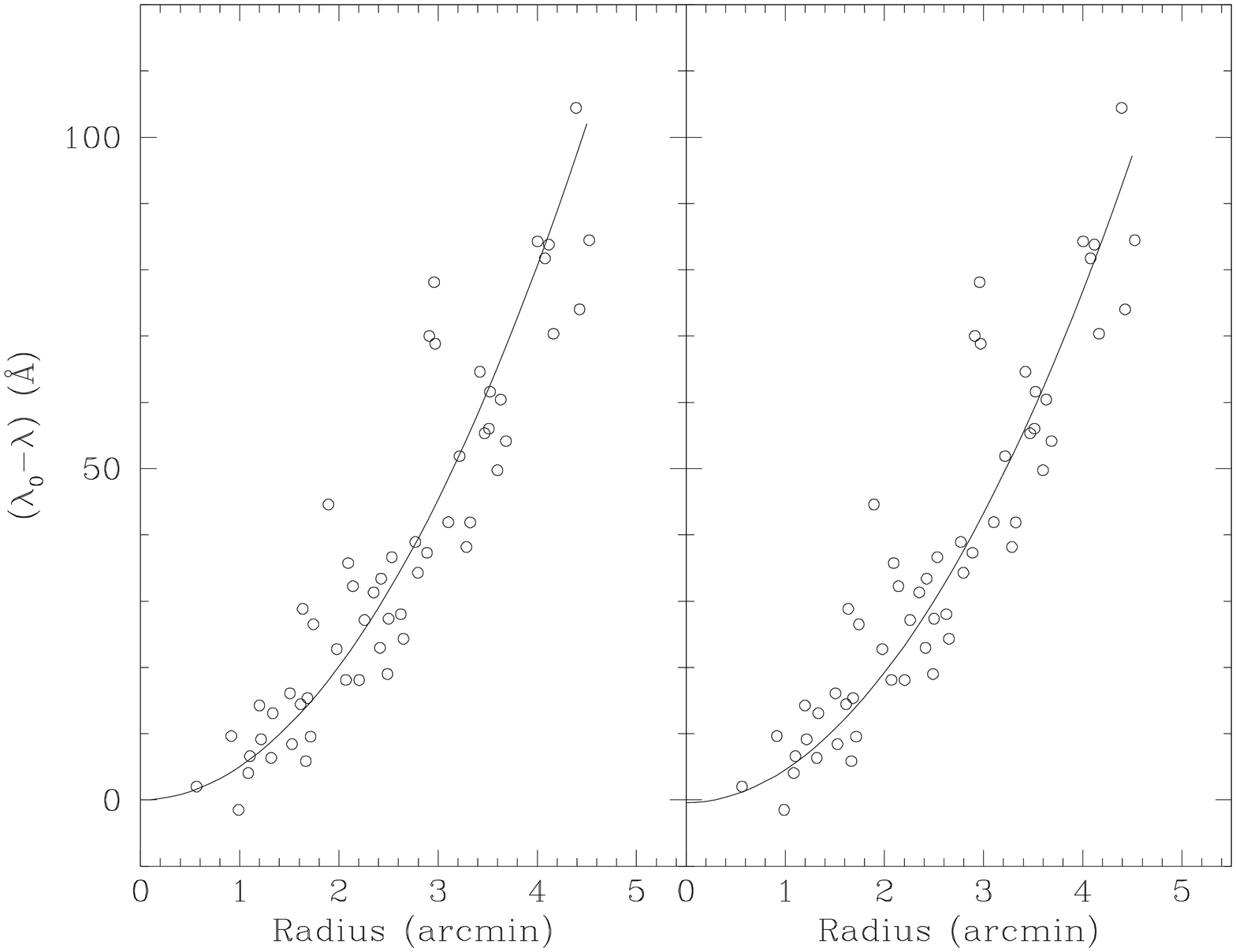}
\caption{$(\lambda_0-\lambda)$ versus radius for the H$\alpha$ line of galaxies in the cluster Cl0024+1654. 
All galaxies with spectroscopic data from \citet{b11} are considered, except some spurious (although few 
spurious are likely still present). Left: using equation (28) results in an overall fitting error of 10.5 \AA. 
Right: using equation (29) results in an overall fitting error of 10.7 \AA. It is difficult choosing one or 
another fitting given the dispersion of the spectroscopic data, although the fitting using equation (29) 
looks better around $\sim 4$  arcmin, as expected.}
\end{figure*}

\section{Conclusions}

The main conclusions found along this work, applied to OSIRIS TFs calibration, can be summarized as follows.

Etalons working at narrow spacings, such as TF, specially those with multilayered thick coatings for working 
at wide spectral ranges, have effective gaps smaller than those given by the nominal plate spacing. Also, at 
non-normal incidence, an anomalous wavelength dependence phase effect, i.e. departure from the theoretical 
dependence (10), is produced. This effect will depend on the coatings used, and must be characterized by 
empirical fitting.

The characterization of the anomalous phase effect of OSIRIS red TF, using spectral calibration lamps, and 
emission line astronomical targets, shows that this effect does not depend on instrument rotation, 
or environmental conditions.

The optical centre of the OSIRIS red TF in the instrument focal plane, expressed in detector coordinates, 
is at: ($-11\pm 1, 976\pm 1$  pix), with respect to ($x,y$) CCD2 reference pixels, and was found stable 
with respect to rotation, environmental conditions, or order of interference.

The observed departures from the theoretical expression (10 or 11) increase not only with radius but with 
wavelength as well, and are well beyond the tuning accuracy even within the monochromatic FOV.

These departures cannot be physically represented by simply changing the constant in equation (10) or (11), 
since the focal lengths of the telescope, collimator or camera do not vary. 

An empirical expression calibrating the wavelength dependence of the red TF across OSIRIS FOV at any 
wavelength, and with an accuracy better than 4\AA\ even at the edge of the OSIRIS red TF FOV is provided. 
For specific observing programs requiring an accuracy of the order of the OSIRIS TF tuning, a second order
iterative expression is provided as well. Similar procedure and analysis will be performed for 
characterizing the wavelength dependence with respect to the distance to the optical centre for the 
OSIRIS blue TF. 

\section*{Acknowledgments}

This work has been partially funded by the Spanish Ministry of Science and Innovation (MICINN) under the Consolider-Ingenio 2010 Program grant CSD2006-00070: First Science with the GTC (http://www.iac.es/consolider-ingenio-gtc), AYA2011-29517-C03-01, and AYA2011-29517-C03-02. Observations presented in this paper were made with the Gran Telescopio Canarias (GTC), installed in the Spanish Observatorio del Roque de los Muchachos of the Instituto de Astrof'sica de Canarias, on the island of La Palma.

\bsp

\label{lastpage}

\end{document}